\documentclass[12pt]{iopart}
\usepackage{times}
\usepackage{graphicx}

\newcommand{\Eins}
           {\;\smash{\raisebox{-0.5ex}{$\!\!\stackrel{\!\mbox{1}
            \hspace{-0.4ex}\rule[0.0ex]{0.06ex}{1.60ex}}{ }$}}}
\begin{document}

\title{Symmetric polynomials in physics}
\author{H J Schmidt and J Schnack}
\address{Department of Physics, University of Osnabr\"uck, Germany}
\begin{abstract}
We give two examples where symmetric polynomials play an important r\^{o}le
in physics: First, the partition functions of ideal quantum gases are closely
related to certain symmetric polynomials, and a part of the corresponding theory
has a thermodynamical interpretation. Further, the same symmetric polynomials also
occur in Berezin's theory of quantization of phase spaces with constant curvature.
\end{abstract}

\section{Introduction}

It often happens that mathematical theories have unexpected applications in physics.
In the present case of the theory of symmetric polynomials (SP) we have, additionally,
the remarkable situation, that physicists have re-discovered certain fragments of the
theory of SP in order to solve problems in few-particle quantum statistical mechanics
\cite{l61},\cite{bf93},\cite{l97},\cite{ss98},\cite{ss99},\cite{acp00}.
Actually, it turns out that a part of the physical theory of ideal quantum gases
is equivalent to a part of the theory of SP if a certain translation scheme is applied,
see below. The central idea of this scheme, namely that partition functions can be
considered as evaluations of certain SP, is not novel, but appeared at various places
in the literature, see \cite{b92},\cite{mss97},\cite{a01},\cite{jpv01}, often in the context
of generalized statistics. Nevertheless,
the relevance for the problems treated in the above-mentioned articles and
the consequences of this observation seem to have remained largely unnoticed.\\

A second field where SP might be important tools is the theory of quantization in the form
suggested by F.~A.~Berezin \cite{b75} and subsequently further developed, see e.~g.~\cite{e02}.
Here the same SP as in quantum statistical mechanics occur in the expansion of the
quantization operator for two-dimensional phase spaces with constant curvature. One
could speculate about the underlying reasons and possible extensions of this connection.

\section{SP and ideal quantum gases}
\label{SPIQG}
We will only explain the basic idea of the connection between SP and quantum
statistical theory of ideal gases. Further details may be found in \cite{ss01} and \cite{b01}
and in the literature quoted there. \\

It is well-known that the eigenstates of the $N$-particle Hamiltonian without
interactions can be characterized by ``occupation number sequences" $i\mapsto n_i$. Here
$n_i$ is the occupation number of the $i$-th energy level $E_i$ of the
$1$-particle Hamiltonian. Hence
\begin{equation}\label{1.1}
\sum_i n_i = N,
\end{equation}
and, for fermions, additionally
\begin{equation}\label{1.2}
 n_i \in \{0,1\}.
\end{equation}
Equivalently, each eigenstate can be characterized by a monomial of degree $N$
\begin{equation}\label{1.3}
{\bi x}^{\bi n}\equiv \prod_i x_i^{n_i},
\end{equation}
where the $x_i$ are abstract, commuting variables corresponding to the
energy levels and ${\bi n}$ denotes the whole occupation number sequence.
The energy eigenvalues corresponding to these eigenstates are
\begin{equation}\label{1.4}
E=\sum_i E_i n_i .
\end{equation}
Now we can express the $N$-particle partition function as
\begin{eqnarray}\label{1.5}
Z_N^\pm (\beta) = \sum_E e^{-\beta E}
&=&
\sum_{\bf n} \exp \left( -\beta \sum_i E_i n_i \right)\\  \label{1.6}
=\sum_{\bf n} \prod_i  \left(e^{ -\beta E_i} \right)^{n_i}
&=&
\left.\underbrace{\sum_{\bf n} \prod_i x_i^{n_i}}\right|_{x_i=\exp(-\beta E_i)}\\
&&
\equiv\left\{
\begin{array}[c]{ll}
b_N(x_1,x_2,\ldots) & \mbox{ : Bosons}(+) \\
f_N(x_1,x_2,\ldots) & \mbox{ : Fermions}(-)
\end{array}
\right.
\end{eqnarray}
Here the sum over ${\bi n}$ is subject to the constraint (\ref{1.1}) for bosons,
and to (\ref{1.1}), (\ref{1.2}) for fermions. Since these constraints are invariant
under permutations of the variables $x_1, x_2,\ldots$, the resulting sum of the
monomials ${\bi x}^{\bi n}$ in (\ref{1.6}) will be a symmetric polynomial of the
$x_1, x_2,\ldots$. We call these SP ``fermi polynomials " $f_N$ or
``bose polynomials" $b_N$, respectively. In the theory of SP the $f_N$ are called
``elementary SP" and the $b_N$ ``complete SP". However, in this article we will
stick to our more physical nomenclature.\\

\begin{figure}
\begin{center}
\includegraphics{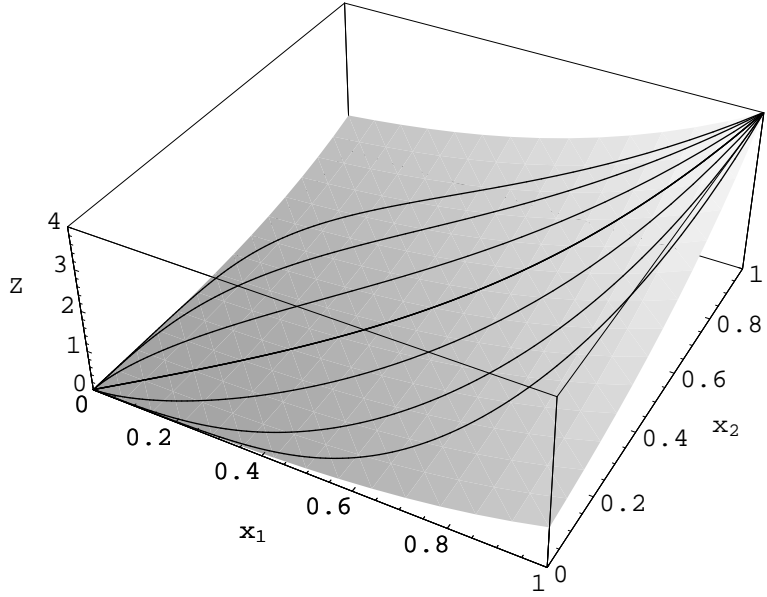}
\end{center}
\caption{This figure shows the graph of the bose ploynomial
$b_3=x_1^3+x_1^2 x_2 + x_1 x_2^2 + x_2^3$
with $N=3$ and $L=2$.
A selected number of curves parametrized by the inverse temperature $\beta$ is
shown which illustrate the partition functions for special systems obtained by the
evaluation of $b_3$ at the values $x_1(\beta)=\exp(-\beta E_1)$ and
$x_2(\beta)=\exp(-\beta E_2)$.}
\label{figure 1}
\end{figure}

The partition function $Z_N(\beta)$ of a particular system is obtained by evaluation
of the corresponding SP along the curve $\beta\mapsto x_i(\beta)=\exp(-\beta E_i)$, see figure 1.
Hence we have a $1:1$ correspondence between certain SP and certain ``partition types"
of ideal gases. Here the ``partition type" of a system is given by the number $N$ of
particles, the number $L\in\{1,2,\ldots,\infty\}$ of abstract energy levels (or the
dimension of the $1$-particle Hilbert space) and the type of the statistics, Bose or
Fermi. The values $E_i$  of the energy levels, including their degeneracy, only determine
the system and its particular partition function. Hence a SP corresponds not to a single
system but to a large class of systems. It is then obvious, that mathematical relations
between the $f_N$ and the $b_N$ can be translated into physical relations between
the corresponding partition functions, irrespective of the values of the $E_i$. \\


There is a third kind of SP with a physical meaning in the theory of ideal gases, the
``power sums"
\begin{equation}\label{1.10}
p_n \equiv \sum_i x_i^n.
\end{equation}
Evaluation at $x_i=\exp(-\beta E_i)$ gives
\begin{equation}\label{1.11}
p_n |_{x_i=\exp(-\beta E_i)}= \sum_i\exp(-n\beta E_i) = Z_1(n\beta).
\end{equation}
One of the central results of the elementary theory of SP is that each of the above
families of SP, the $f_N$, the $b_N$, and the $p_n$, can be used as a ``basis" of SP,
in the sense that any SP can be expressed as a polynomial of the $f_N$ (resp.~$b_N$ or
$p_n$). This implies that the fermionic partition functions can be expressed by means
of the bosonic ones and vice versa. Moreover, both in turn can be expressed
by means of the $1$-particle partition function at different inverse temperatures
$n \beta$. These relations are more or less known in the physical literature but their
origin in the theory of SP has only recently be disclosed \cite{ss01}.\\

SP can be defined through their generating functions. In statistical mechanics the corresponding
generationg functions are called ``grand canonical partitions functions" and will be denoted
by $B(z)$ for bosons and $F(z)$ for fermions. The formal parameter $z$ is physically interpreted
as the fugacity $z=\exp(-\beta \mu)$, where $\mu$ is the chemical potential. The physical domain
of $z$ is $(0,1)$ for bosons and $(0,\infty)$ for fermions.\\

There exists a fundamental symmetry $\omega:\Lambda\longrightarrow\Lambda$
of the ring $\Lambda$ of SP, which maps the $f_N$ onto the $b_N$, see \cite{m79}.
It is connected with the equation $F(z) B(-z)=1$. The physical interpretation of this relation is that the
fermionic grand canonical partitions function is related to the analytical continuation
of the bosonic one to negative $z$ and vice versa. Similar relations for other thermodynamic functions
are implied. Whether analytical continuation is possible depends on the system under consideration.
For the system of particles in a box it has been shown that analytical continuation of the partition
functions is possible in the thermodynamic limit \cite{l97}
and the resulting Bose-Fermi symmetry has been discussed. Another Bose-Fermi symmetry  has been
observed for particles in odd space dimensions confined by a common harmonic oscillator potential
\cite{ss99}. Here $Z_1(\beta)=-Z_1(-\beta)$ implies $Z_N^{+}(\beta)=(-1)^N Z_N^{-}(-\beta)$.
Since here analytical continuation is involved w.~r.~t.~the $\beta$-plane, it is not yet clear
how this symmetry is related to that given by $\omega$\\

These and further relations and translation schemes will be sketched in the table below.
For details see \cite{ss01} and, as a standard reference for the theory of SP,
\cite{m79}.

\begin{tabular}{l|l}
{\it Physics} & {\it Mathematics}\\
&\\
\hline
&\\
\rule{0mm}{10mm}Abstract energy levels & Variables $x_1, x_2, \ldots$\\
&\\
\rule{0mm}{10mm}Occupation by $N$ particles & Monomials $ {\bf x}^{\bf n}$\\
&\\
\rule{0mm}{10mm}{Partition types}
&
\rule{0mm}{10mm}Symmetric polynomials $p({\bf x}) = \sum_{\bf n} {\bf x}^{\bf n}$\\
&\\
\rule{0mm}{10mm} -
&
\rule{0mm}{10mm}Ring of symmetric polynomials $\Lambda$\\
&\\
\rule{0mm}{10mm}Partition function $Z_N(\beta)$
&
\rule{0mm}{10mm}Evaluation of $p({\bf x})$ at $x_i= e^{-\beta E_i},\; i=1,\ldots, L$ \\
\rule{0mm}{10mm}
$
\left. \begin{array}{l}F(z)\\
B(z)\end{array}
\right\}
$
\parbox[c]{4cm}{Grand canonical partition functions}
& \rule{0mm}{10mm}
$
\left. \begin{array}{l}F(z)
\\
B(z)
\end{array}
\right\}
$
\parbox[c]{5cm}{Generating functions for
$f_n,\, b_n$}\\
&\\
\rule{0mm}{10mm}Fugacity $z=e^{-\beta \mu}\;$&
$F(z)
\equiv \prod_{i=1}^{L}{(1+x_i\, z)}
=
\sum_{n=0}^{L}{f_n\, z^n}$
\\
&
$
B(z)
\equiv \prod_{i=1}^{L}{\frac{1}{1-x_i\, z}}
=
\sum_{n=0}^{L}{b_n\, z^n} $
\\
\rule{0mm}{10mm}\parbox[l]{6cm}{Bose-
Fermi symmetry by analytical continuation
in the $z$-plane \cite{l97}}
&
\rule{0mm}{10mm}$B(z)F(-z)=1$\\
&
\rule{0mm}{10mm}
$
\left. \begin{array}{l}\omega:\Lambda \longrightarrow \Lambda
\\
\omega:f_n\mapsto b_n
\end{array}
\right\}
$
\parbox[c]{5cm}{involutive automorphism of graded rings}\\
&\\
\rule{0mm}{10mm}$\sum_i\left( e^{-\beta E_i}\right)^n = Z_1(n \beta)$
&
$p_n,$\mbox{  evaluated}
\\
&\\
-
&
Generating function:\\
&
$P(z)=\frac{d}{dz} \log B(z) = \sum_{n=1}^L p_n z^n$\\

\rule{0mm}{10mm}$\langle N \rangle$
&
\rule{0mm}{10mm}$ z P(z) $  \\
&\\
\rule{0mm}{10mm}
\parbox{6cm}{
Landsberg's identities (Appendix E of his 1961 textbook,
rediscovered several times)
}
&
Newton's identity:
$n f_n =
\sum_{r=1}^n (-1)^{r-1} p_r f_{n-r}
$
\\
&\\
\rule{0mm}{10mm}   &
\end{tabular}

\section{SP and Berezin quantization}
The main idea of Berezin's approach to quantization is the use of
``generalized coherent states" (GCS) in order to establish an
approximate equivalence between  classical and quantum observables and states.
GCS will be denoted by $|\alpha\rangle$ where the ``parameter"
$\alpha$ runs through some phase space $M$. The most important property of GCS
is the completeness relation
\begin{equation}\label{2.1}
\Eins = \int_M |\alpha\rangle\langle \alpha|\; d\alpha.
\end{equation}
CGS are used to map operators $A$ onto functions on $M$ by means of
\begin{equation}\label{2.2}
j(A)(\alpha) \equiv\frac{\langle\alpha|A|\alpha\rangle}{\langle\alpha|\alpha\rangle}.
\end{equation}
The adjoint map is given by
\begin{equation}\label{2.3}
j^*(f)= \int_M f(\alpha)|\alpha\rangle\langle \alpha|\; d\alpha,
\end{equation}
where we ignore all questions about the exact domains of definition
of (\ref{2.2}) and (\ref{2.3}). The ``quantization operator"
$j\circ j^\ast$ maps functions onto functions. It is used for
semiclassical expansions of physical quantities
w.~r.~t.~powers of $h$, where $h$ is a formal
parameter (``Planck's constant") on which the Hilbert space and the GCS
depend.\\

In the case of a flat phase space $M={\bf R}^2$, the GCS are chosen as the usual
coherent states
discovered by E.~Schr\"odinger. Then
\begin{equation}\label{2.4}
j\circ j^\ast = \exp (-h^2 \Delta),
\end{equation}
where $\Delta$ is the Laplace operator in ${\bf R}^2$, see e.~g.~\cite{ks85}.
There are at least two other cases where $j\circ j^\ast$ can be calculated: The
Lobachevskij plane $L$ and the $2$-sphere $S^2$. Here $j\circ j^\ast$ is given in the form
of an infinite product, see \cite{b75}, which resembles the grand canonical
partition functions considered in the previous section. Hence   $j\circ j^\ast$
can be expanded into power series w.~r.~t.~$\Delta$, where the coefficients are
bose polynomials for $L$ and fermi polynomial for $S^2$. $\Delta$ is the corresponding
Laplace-Beltrami operator of $L$, resp.~$S^2$. More explicitely:\\

Case 1
\begin{eqnarray}  \nonumber
M
&=&
\mbox{{\sc Lobachevskij} plane } L
\\     \nonumber
|\alpha \rangle
&:&
\mbox{{\sc Berezin's} coherent states }\cite{b75}\\
j\circ j^*
&=&
\prod_{n=0}^\infty
\left(
1-\frac{h^2}{(1+nh)(1-(n-1)h)}\Delta
\right)^{-1} \\
&=&
\sum_{N=0}^\infty b_N(x_1,x_2,\ldots) \Delta^N\\
\mbox{where}
&&
x_n=\frac{h^2}{(1+nh)(1+(n-1)h)}
\end{eqnarray}

Case2

\begin{eqnarray} \nonumber
M
&=&
\mbox{$2$-sphere } S^2
\\    \nonumber
|\alpha \rangle
&:&
\mbox{{\sc Bloch's} coherent states } \cite{ks85}\\
h
&=&
\frac{1}{2s}\quad (s:\mbox{ spin quantum number})\\
j\circ j^*
&=&
\prod_{n=0}^\infty
\left(
1+\frac{h^2}{(1+nh)(1+(n+1)h)}\Delta
\right) \\
&=&
\sum_{N=0}^\infty f_N(x_1,x_2,\ldots) \Delta^N\\
\mbox{where}
&&
x_n=\frac{1}{(2s+n)(2s+n+1)}
\end{eqnarray}

It is remarkable, that the bose and fermi polynomials occur for the two simplest
cases of $2$-dimensional phase spaces with constant negative and positive curvature.
Both kind of polynomials can be considered as special cases of the so-called Schur
polynomials $s_\lambda$, where $\lambda$ denotes a Young diagram or, equivalently,
a partition of $N$, see \cite{m79}. In fact, $b_N=s_{(N)}$ and $f_N= s_{(1^N)}$, where
$(N)$ (resp.~$(1^N)$) is the Young diagram consisting of a single row (resp.~column).
One might wonder whether also other Schur polynomials occur in the expansion of
the quantization operator for other choices of phase spaces and GCS.


\end{document}